\documentclass[11pt]{article}

\usepackage[dvips]{graphicx}

\newcommand{\be}{\begin{equation}}
\newcommand{\ee}{\end{equation}}
\newcommand{\ra}{\rightarrow}
\newcommand{\wt}{\widetilde}
\newcommand{\si}{\phi_{\infty}}

\newcommand{\smfrac}[2]{{\textstyle{#1\over#2}}}
\def\half{\smfrac{1}{2}}

\begin{document}
\title{Neutron Stars in a Varying Speed of Light Theory}

\author{A. W. Whinnett\thanks{Email: visitor2@pact.cpes.susx.ac.uk}\\
        Astronomy Centre,\\ University of Sussex, \\ Falmer,
        \\ Brighton, \\ BN1 9QJ, \\ United Kingdom}

\maketitle

\begin{abstract}

We study neutron stars in a varying speed of light (VSL) theory of
gravity in which the local speed of light depends upon the value of a
scalar field $\phi$. We find that the masses and radii of the stars 
are strongly dependent on the strength of the coupling between $\phi$
and the matter field and that for certain choices of coupling
parameters, the maximum neutron star mass can be arbitrarily small. We
also discuss the phenomenon of cosmological evolution of VSL stars
(analogous to the gravitational evolution in scalar-tensor theories) 
and we derive a relation showing how the fractional change in the energy
of a star is related to the change in the cosmological value of the
scalar field.

\end{abstract}

PACS numbers: 04.40.Dg, 04.50.+h

%%%%%%%%%%%%%%%%%%%%%%%%%%%%%%%%%%%%%%%%%%%%%%%%%%%%%%%%%
\section{Introduction}\label{INT}
%%%%%%%%%%%%%%%%%%%%%%%%%%%%%%%%%%%%%%%%%%%%%%%%%%%%%%%%%

Recently, there has been interest in the possibility that the speed of
light $c$ might have been larger in the past. The primary theoretical
reason for considering this possibility is that such a variation in
$c$ could solve the horizon and flatness problems of big-bang
cosmology, without needing to postulate the existence of an inflationary
epoch in the early history of the Universe \cite{Barrow}. This purely
theoretical work was given added impetus more recently after the
discovery by Webb {\it et al}. \cite{Webb} that, according to
observations of wavelength shifts in
the absorption lines of distant quasars, the fine structure constant 
$\alpha:=e^{2}/(\hbar c)$ seems to have been smaller in the past. This is
consistent with the assumption  that $c$ was larger in the past and,
although more recent evidence \cite{Martins}
has constrained $\alpha$ variations more
strongly than the Webb data, the possibility of a time
varying $\alpha$ remains. 

One can assume that any changes in the value
of $\alpha$ are due to variations in one or both of $\hbar$ and
$e$. However, the physical consequences of allowing either of these to
vary are different from those that arise from a varying $c$
theory. For example, Avelino \& Martins \cite{Avel} have shown that
the horizon and flatness problems can only be solved by assuming that
$c$ is varying. If instead one assumes that either of the other two
constants is responsible for the variation of $\alpha$, the horizon
and flatness problems remain.

To analyse the consequences of allowing a physical ``constant'' to vary,
one needs a self consistent theory which is capable of describing this
variation and its effect on other physical processes.
Since $\alpha$, and by implication $c$, may change significantly only over
cosmological time and distance scales, it is necessary to develop a
theory of gravity which embodies such behaviour. This causes 
immediate problems when trying to formulate a
varying speed of light (VSL) theory since the constancy of $c$ is 
a fundamental assumption of relativistic physics. Several
attempts have been made to reconcile such a variation with a
relativistic gravitational theory, including a bi-metric 
theory developed by
Clayton \& Moffat \cite{Moffat} and Magueijo's  scalar field theory
\cite{Mag1,Mag2}. It is the latter that we shall focus on here.

In Magueijo's theory, the timelike coordinate in a coordinate chart no
longer represents time as measured by a physical process and is
instead replaced by a fourth length coordinate
$\zeta$. When described in terms of this new coordinate, the structure
and properties of a manifold are the same as in any relativistic
theory of gravity and all of the rules of differential geometry remain
unchanged. In particular, the manifold has a metric, a Riemann
curvature tensor, and so on, all of whose components can be expressed
in a $(\zeta,x^{i})$ coordinate basis, where the $x^{i}$ are spatial
coordinates. Time as measured by physical processes is not in general
usable as a coordinate on the manifold. However, once a solution is
found, physically measured time intervals can then be determined using
the relation $dt=c^{-1}d\zeta$, where $c$ is treated as a scalar field
on the manifold. The fact that $t$ is no longer necessarily
a timelike coordinate is apparent when one considers
the possibility that in a solution, surfaces of constant $c$ are not
necessarily spacelike, so that 
$\nabla_{a}c$ will not be timelike. The VSL theory is
superficially similar to Brans Dicke (BD) theory 
except that the scalar field,
instead of determining the strength of the gravitational field,
governs the rate of physical processes and, by
implication, the rate at which clocks record time.

In any particular VSL theory, one would naturally expect spatial
as well as temporal variations in $c$ in a spacetime with an
inhomogeneous matter field.
This is indeed the case, and
Magueijo has analysed vacuum, spherically symmetric solutions
representing both the weak field produced far from a central source 
and VSL black holes \cite{Mag2}.
In this paper, we analyse the structure of neutron stars in Magueijo's
VSL theory. The purpose of this is twofold: first, to determine the
degree to which a VSL neutron star differs from its counterpart in general
relativity (GR) or BD theory and second, to serve as a starting
point for the theoretical treatment of binary pulsars in a VSL theory,
with the ultimate aim of using binary pulsar data to constrain the
parameters of the theory.

The plan of the paper is as follows. In Section \ref{ACT} we give the
action for the theory and derive the 
field equations for spherically symmetric spacetimes. In Section
\ref{EXT} we discuss the vacuum, exterior part of the solutions. In
Section \ref{STATE} we give the equation of state (EOS) of the fluid
interior and discuss the issue of particle conservation, while in
Section \ref{MASS} we find the expression which describes the total
energy of the solutions. In Section \ref{NUM} we present the results of
numerical integrations of the equations of structure and in Section
\ref{SENS} we discuss the evolution of the mass of a star in a
cosmological setting. Finally, in Section \ref{CONC} we present a few
concluding remarks.

%%%%%%%%%%%%%%%%%%%%%%%%%%%%%%%%%%%%%%%%%%%%%%%%%%%%%%%%%
\section{Action and Field Equations}\label{ACT}
%%%%%%%%%%%%%%%%%%%%%%%%%%%%%%%%%%%%%%%%%%%%%%%%%%%%%%%%%

In Magueijo's VSL theory, the local speed of light $c$ is determined by
the value of a scalar field $\phi$ such that $c=c_{\infty}e^{\phi}$, where
$c_{\infty}$ is the speed of light measured far from all gravitating
sources. In units in which $G=1$, the action
is given by\footnote{The constants $\epsilon$ and $\beta$ we use here
correspond to the constants $a$ and $b$ used by Magueijo
\cite{Mag1,Mag2}. We use the former notation to avoid confusion with
tensor indices.} 
\be\label{action}
   I=\frac{1}{16\pi}\int_{\cal V} d^{4}x\sqrt{-g}\left[e^{\epsilon\phi}
   \left({\cal R}-\kappa g^{ab}
   \nabla_{a}\phi\nabla_{b}\phi\right)
   +16\pi e^{\beta\phi}L_{m}\right]
   -\frac{1}{8\pi}\int_{\partial\cal V}d^{3}x\sqrt{h}\;
   e^{\epsilon\phi}\Theta,
\ee
where ${\cal V}$ is an arbitrary volume of integration, 
$\partial{\cal V}$ its closed boundary, and $h_{ab}$ and $\Theta$ the induced
metric and extrinsic curvature on this boundary. The surface term 
guarantees that the variational derivative of $I$
vanishes on $\partial {\cal V}$ and is
important when determining the mass of an asymptotically flat
spacetime. The constants $\epsilon$, $\beta$ and $\kappa$ are arbitrary.
In \cite{Mag1} the condition $\epsilon-\beta=4$ was imposed since the
units of the matter and gravitational terms of the 
action (\ref{action}) differ by a factor of $c^{4}$. We shall assume
this condition holds here. In addition, we shall
assume that the ``minimum coupling'' assumption of {\cite{Mag1} holds
so that $L_{m}$ is independent of $\phi$.

The generalised Einstein equations which lead from this action are
\be\label{genEin}
   G_{ab}=\left(\kappa+\epsilon^{2}\right)\nabla_{a}\phi\nabla_{b}\phi
   -(\half\kappa+\epsilon^{2}) g_{ab}\nabla^{c}\phi\nabla_{c}\phi
   +\epsilon(\nabla_{a}\nabla_{b}\phi-g_{ab}\nabla^{c}\nabla_{c}\phi)
   +8\pi e^{(\beta-\epsilon)\phi}T_{ab},
\ee
where $T_{ab}$ is the matter energy-momentum tensor,
while the scalar field satisfies the wave equation
\be\label{boxS}
   \nabla^{a}\nabla_{a}\phi=-\epsilon\nabla^{a}\phi\nabla_{a}\phi
   +\frac{8\pi G e^{(\beta-\epsilon)\phi}}{2\kappa+3\epsilon^{2}}\left(
   \epsilon T-2\beta L_{m}\right).
\ee
The local matter conservation law for the theory is
\be\label{Matter}
   \nabla_{a}T^{ab}=-\beta\left(T^{ab}-g^{ab}L_{m}\right)\nabla_{a}\phi.
\ee
In the vacuum case, where $L_{m}=0$ and $T_{ab}$ vanishes, the action
and field equations are identical to BD theory with
\be
   \omega=\frac{\kappa}{\epsilon^{2}},\;\;\;\;\Phi=e^{\epsilon\phi},
\ee
where $\omega$ and $\Phi$ are the BD coupling parameter and scalar
field. 

We are interested here in asymptotically flat, static and spherically
symmetric solutions to the VSL field equations.  
We use isotropic coordinates and write the line element as
\be
   ds^{2}=-B\,d\zeta^{2}+A\left(dr^{2}+r^{2}\,d\theta^{2}
   +r^{2}\sin^{2}\theta\right),
\ee
where $A$ and $B$ are functions of $r$, and $\zeta$ is a timelike
coordinate with the units of length. We assume that the matter is
composed of a perfect fluid with pressure $p$, energy density
$\rho$ and a fluid 4-velocity $U^{a}$, for which the 
energy-momentum tensor has the usual form
\be\label{EMtensor}
   T^{ab}=(\rho+p)U^{a}U^{b}+p g^{ab}.
\ee
In the coordinates we are using, $U^{a}=(B^{-1/2},0,0,0)$ and is
orthogonal to the hypersurfaces of constant $\zeta$.

To integrate the resulting
field equations, it is convenient to introduce an auxiliary function 
\be\label{defz}
   Z:=\frac{1}{2B}\frac{dB}{dr}+\frac{1}{2A}\frac{dA}{dr}.
\ee
In terms of these variables, the field equations reduce to
the following set of coupled ODEs:
\be\label{ddAdrr}
   A^{\prime\prime}=\frac{3A^{\prime 2}}{4A}
   -\frac{2A^{\prime}}{r}+\epsilon\phi^{\prime}
   \left(AZ-\frac{A^{\prime}}{2}\right)
   -\frac{A\kappa\phi^{\prime 2}}{2}+8\pi A^{2}e^{(\epsilon-\beta)\phi}\left[
   \frac{\rho\left(\epsilon^{2}-\epsilon\beta\right)+3p\epsilon^{2}}{2\kappa
   +3\epsilon^{2}}-\rho\right],
\ee
\be\label{dZdr}
   Z^{\prime}=-\frac{3Z}{r}-Z^{2}-\epsilon\phi^{\prime}
   \left(Z+\frac{1}{r}\right)+8\pi Ae^{(\epsilon-\beta)\phi}
   \left[2p+\frac{\rho \epsilon(\epsilon-2\beta)-3p\epsilon^{2}}{2\kappa
   +3\epsilon^{2}}\right],
\ee
\be\label{ddSdrr}
   \phi^{\prime\prime}=-\phi^{\prime}\left(\epsilon\phi^{\prime}
   +Z+\frac{2}{r}\right)+8\pi Ae^{(\epsilon-\beta)\phi}\left[\frac{3p\epsilon
   -\rho(\epsilon+2\beta)}{2\kappa+3\epsilon^{2}}\right]
\ee
and
\be\label{dpdr}
   p^{\prime}=-(\rho+p)\left(Z-\frac{A^{\prime}}{2A}
   +\beta\phi^{\prime}\right),
\ee
where the prime denotes $\frac{d\;}{dr}$. 

%%%%%%%%%%%%%%%%%%%%%%%%%%%%%%%%%%%%%%%%%%%%%%%%%%%%%%%%%
\section{Exterior Solution, Boundary Values and Scaling 
Relations}\label{EXT}
%%%%%%%%%%%%%%%%%%%%%%%%%%%%%%%%%%%%%%%%%%%%%%%%%%%%%%%%%

For a given set of values of $\epsilon$, $\beta$ and $\kappa$, the 
solutions are parameterised by the central density $\rho_{0}$. As in
GR, the surface of a stellar solution is defined to be where the
pressure vanishes and at this point the solution is matched to a
vacuum exterior. Since the vacuum field equations for this theory are
identical to BD theory, the vacuum solutions are also
identical and, in terms of the variables we are using here, are given
by
\be\label{AA}
   A=A_{\infty}\left(1+\frac{\mu}{r}\right)^{4}
   \left(\frac{r-\mu}{r+\mu}\right)^{2(1-C/\lambda-1/\lambda)},
\ee
\be\label{ZZ}
   Z=\frac{2\mu}{\mu^{2}-r^{2}}
\ee
and
\be\label{SS}
   \phi=\phi_{\infty}+\frac{C}{\epsilon\lambda}
   \log\left(\frac{r-\mu}{r+\mu}\right),
\ee 
where the subscript ``$\infty$'' denotes the value of quantities at
spacelike infinity. In these equations $\mu$, $C$ and $\lambda$ 
are constants and satisfy the constraint
\be\label{lambda}
   \lambda^{2}=(C+1)^{2}
   -C\left(1-\frac{\kappa C}{2\epsilon^{2}}\right).
\ee
Equations (\ref{AA}) and (\ref{ZZ}) together give the more familiar
metric component
\be\label{BB}
   B=B_{\infty}\left(\frac{r-\mu}{r+\mu}\right)^{2/\lambda}.
\ee
These equations hold only when $\epsilon\neq 0$.

We shall assume that the interior neutron star solution matches
smoothly to the above vacuum solution at the boundary of the star, in
the sense that $A$, $Z$, $\phi$ and the derivatives of these functions
are continuous at the boundary. We can then find the values of the
parameters $\mu$, $C$ and $\lambda$ as follows. Differentiating eqns
(\ref{AA}) to (\ref{SS}), one can show that
\be
   X:=Z+\epsilon\phi^{\prime}=\frac{2\mu^{2}}{r^{3}-\mu^{2}r}.
\ee
Rearranging this equation, we  have
\be\label{mucalc}
   \mu=\sqrt{\frac{Xr^{3}}{4+Xr}},
\ee
which allows us to determine the value of $\mu$. From eqn
(\ref{ZZ}), one can show that
\be
   \lambda=\left(Z-\frac{A^{\prime}}{2A}\right)
   \frac{2\mu}{r^{2}-\mu^{2}},
\ee
which allows us to calculate $\lambda$ once $\mu $ has been found.
Finally, from eqn (\ref{SS}), we have
\be\label{ccalc}
   C=\frac{\lambda(r^{2}-\mu^{2})\epsilon\phi^{\prime}}{2\mu},
\ee
which allows us to calculate $C$.

The minimal VSL theory mentioned in \cite{Mag2} is the one with
$\epsilon=0$ and its field equations are much simpler than in the 
more general theory. For this particular case, the above
equations have to be
modified. The functions $A$, $B$ and $Z$ are given by eqns (\ref{AA}),
(\ref{ZZ}) and (\ref{BB}) with the constant $C=0$. The scalar field is
now given by
\be\label{SS2}
   \phi=q\log\left(\frac{r^{2}-\mu^{2}}{r^{2}+\mu^{2}}\right),
\ee
where $q$ is a new constant, related to $\lambda$ and $\kappa$ by
\be
   q^{2}\kappa=2-\frac{2}{\lambda^{2}}.
\ee
The Schwarzschild limit discussed in \cite{Mag2} is the
special case of these equations in which $\lambda=1$. Note that
in all cases in which $\epsilon=0$, the $\phi$ field need not be
homogeneous.

Magueijo \cite{Mag2} has shown that, by carrying  
out a weak field analysis and interpreting
the results within the framework of the PPN formalism, 
the parameters  $\kappa$, $\epsilon$ and $\beta$ are related by the
equation 
\be\label{ppnpars}
   \gamma=\frac{\epsilon^{2}+\kappa+\half\epsilon\beta}
   {2\epsilon^{2}+\kappa-\half\epsilon\beta},
\ee
where $\gamma$ is the PPN parameter which governs, amongst other
things, the deflection of light rays near to a massive body such as
the Sun. Its limiting value is currently \cite{Glimit}
\be
   |\gamma-1|\le 0.0003.
\ee
Solving this eqn (\ref{ppnpars}) for $\kappa$ we have
\be\label{ppnpars2}
   \kappa=\frac{\epsilon(\epsilon-\beta)}{\gamma-1}+\half\epsilon\beta
   -2\epsilon^{2},
\ee
which shows that $\kappa$ decreases as $\epsilon\ra 0$ for the case 
when $\epsilon-\beta=4$. 
For values of $\epsilon$ of the order unity, eqn
(\ref{ppnpars2}) restricts $\kappa$ to have a large value
similar to that of $\omega$ in BD theory. However, if we allow
$\epsilon$ to have an arbitrarily small value, there is no minimum value
to $\kappa$.

%%%%%%%%%%%%%%%%%%%%%%%%%%%%%%%%%%%%%%%%%%%%%%%%%%%%%%%%%
\section{Equation of State and Particle Conservation}\label{STATE}
%%%%%%%%%%%%%%%%%%%%%%%%%%%%%%%%%%%%%%%%%%%%%%%%%%%%%%%%%

We assume that the perfect fluid matter of the star consists solely of
neutrons, each with rest mass $m$ and local number
density $n$. Denoting the internal energy per particle by $\Pi$, the
matter Lagrangian is then given by
\be\label{Lagdef}
   L_{m}=-mc^{2}n(1+\Pi)=-\rho,
\ee
the latter equality defining the relationship between $n$ and $\rho$.
Under the minimum coupling assumption outlined in \cite{Mag1}, the
particle rest energy $mc^{2}$ is constant and hence $L_{m}$ is
independent of $\phi$. 

We shall consider two equations of state. The first is that of a
a non-interacting gas of Fermions, which may be given in
the parametric form \cite{LanLif}
\begin{eqnarray}\label{EOS1}
   p=\frac{m^{4}c^{5}}{24\pi^{2}\hbar^{3}}\left[\chi(2\chi^{2}-3)
   \sqrt{\chi^{2}+1}+3\log(\chi+\sqrt{\chi^{2}+1})\right] \nonumber\\
                                                                   \\
   \rho=\frac{m^{4}c^{5}}{3\pi^{2}\hbar^{3}}
   \chi^{3}\sqrt{\chi^{2}+1}-p.\nonumber
\end{eqnarray}
This means that, in practice, eqn (\ref{dpdr}) becomes a differential
equation for $\chi$. Note that, under the minimum coupling assumption
of \cite{Mag1}, both $\hbar$ and the product $mc$ scale with $\phi$ in the
same way, so that $mc/\hbar$ is independent of $\phi$ and hence so are
numerical coefficients of eqns (\ref{EOS1}).

More realistic equations of state are generally given in tabulated 
form and require complex calculations which take into account particle
phase transitions at various densities. However, 
polytropic equations of state fitted to the tabulated data do exist
and our second EOS is of this type.
The pressure and density are given in the parameterised form
\begin{eqnarray}\label{EOS2}
   p= Kn_{0}\;m\left(\frac{n}{n_{0}}\right)^{\Gamma}\nonumber \\
                                                             \\
   \rho=nm+\frac{Kn_{0}\;m}{\Gamma-1}
   \left(\frac{n}{n_{0}}\right)^{\Gamma},\nonumber
\end{eqnarray}
where the constants are chosen to have the values $n_{0}=1.0\times
10^{44}$m$^{-3}$, $K=0.0195$ and $\Gamma=2.34$. The equations are
parameterised by the particle number density $n$. 
This is the EOS used by Damour and Esposito-Farese in their
investigation of strong field scalar effects in neutron stars
\cite{DEF} and is based on tabulated data given in \cite{Diaz}. 
As before, we assume that the minimum coupling assumption holds so 
these parameters are independent of the local value of $c$.

The first of our equations of state is far from 
realistic. However, we are
interested here primarily in the effect a VSL theory has on the
properties of a neutron star and we expect the main features of
the solutions to be independent of the choice of EOS. In
addition, our choice makes the results we find here
easier to compare with the
GR stars studied by Oppenheimer \& Volkoff \cite{OV} and the BD
solutions of Salmona \cite{Sal}. The second EOS, being
more realistic, will be used to place tentative constraints on the
theory. 

Despite the presence of the factor of $e^{\beta\phi}$ in the matter
part of the action (\ref{action}), the conserved particle number $N$ for
a neutron star solution is the same as in a metric theory, namely
\be\label{Number}
   N=\int_{0}^{r_{s}}4\pi r^{2}A^{3/2}n \;dr,
\ee
where $n$ is the local particle number density. In the second EOS
outlined above, this is the same quantity which parameterises
those equations while in the simpler, non-interacting Fermion EOS,
$n$ and $\chi$ are related by 
\be
   n=\left(\frac{m^{3}c^{3}}{3\pi^{2}\hbar^{3}}\right)\chi^{3}.
\ee
In this relation, the bracketed combination of parameters
is independent of $c$ due to the minimal coupling assumption.

We note here that for a non-perfect fluid matter source,
the conserved charge should in general 
include explicit reference to the fact that $L_{m}$ is not
universally coupled.
One can show that, for matter Lagrangians $L_{m}(\psi)$ dependent on
one or more matter fields $\psi$ which are invariant under a field
transformation $\psi\ra\psi+\delta\psi$,
the N\"{o}ether current associated with the symmetry is given by
\be
   J^{a}=e^{\beta\phi}I^{a},
\ee
where $I^{a}$ is the corresponding current in a metric theory in which
$L_{m}$ is universally coupled to the metric. This means that one
should include a factor of $e^{\beta\phi}$ in the expression for the
conserved charge. Such a factor
appearing in the N\"{o}ether current would
indicate a violation of a local energy conservation law. One
can show that this does not occur for a perfect fluid source as
follows. Dropping the assumption that the solutions are static and
contracting eqn (\ref{Matter}) with $U_{b}$ we have
\be
   U_{b}\nabla_{a}T^{ab}=-\beta U_{b}
   \left[(\rho+p)U^{a}U^{b}+(p+\rho) g^{ab}\right]\nabla_{a}\phi
\ee
which vanishes. Hence, at least for a perfect fluid, the non-spatial
components of eqn (\ref{Matter}) are trivially zero and the right hand
term contributes only to the pressure gradient. Thus there is no
violation of the local energy conservation law in this case.

%%%%%%%%%%%%%%%%%%%%%%%%%%%%%%%%%%%%%%%%%%%%%%%%%%%%%%%%%
\section{Mass}\label{MASS}
%%%%%%%%%%%%%%%%%%%%%%%%%%%%%%%%%%%%%%%%%%%%%%%%%%%%%%%%%

Since the solutions are spherically symmetric and asymptotically flat,
one may write the ADM mass as the integral
\be\label{MassInt}
   M_{ADM}=\int_{0}^{\infty}4\pi r^{2}G_{ab}U^{a}U^{b}\;dr.
\ee
This quantity is independent of the theory of gravity being
considered and its value is determined solely by the metric. 
For both BD and VSL theories, from eqn (\ref{AA}), the ADM mass is
also given by
\be\label{MADM}
   M_{ADM}=\frac{2(C+1)\mu}{\lambda}.
\ee
The Poisson mass, discussed in \cite{Mag2},
may be defined by the relation
\be\label{MPdef}
   M_{P}:=\lim_{r\to\infty}\left(\half r^{2} B^{\prime}\right)
   =\frac{2\mu}{\lambda},
\ee
where the second equality follows from using eqn (\ref{BB}). 
  
In theories of gravity other that GR, $M_{ADM}$ does not
describe the total, physical energy of an asymptotically flat
solution. For example, in BD theory
the energy of a spherically symmetric solution is the Tensor
mass \cite{Lee}
\be\label{MTdef}
   {\widetilde M}_{T}:=\frac{1}{\Phi_{\infty}}\left[M_{ADM}
   -\half\lim_{r\to\infty}\left(r^{2}\frac{d\Phi}{dr}\right)\right]
   =\frac{1}{\Phi_{\infty}}\left(M_{ADM}-\half{\widetilde Q}_{S}\right)
\ee   
so called because it is the active gravitational mass which affects
the motion of a test particle composed entirely of gravitational field
self energy (such as a black hole). The second equality in eqn
(\ref{MTdef}) defines the scalar charge ${\widetilde Q}_{S}$
associated with $\Phi$, which characterises the extent to
which the gravitational energy of the star differs from that of a
similar star in GR, and we use a tilde to denote quantities relating
to BD theory. Equation (\ref{MTdef}) includes an overall factor
of $\Phi_{\infty}^{-1}$ to account for the fact that, as $\Phi$
varies, the strength of the gravitational coupling $G$ varies and the
gravitational energy, which includes implicitly a factor of $G$,
should vary in the same way.

To find an expression for the energy of a solution of the VSL theory, 
we shall find an expression for the Hamiltonian. 
The procedure is almost identical to that given
for dilaton theories by Ho, Kim and Park \cite{HKP}, so we only give
the result here. We consider only a static spacetime; the
generalisation to a time-dependent spacetime is as straight forwards
as it is in GR. We assume that the spacetime is asymptotically flat and use
the usual 3+1 splitting, denoting spacelike hypersurfaces by $\Sigma$
and the hypersurface orthogonal unit timelike vectors by $U^{a}$. The
timelike, 2-dimensional boundary of integration at spacelike infinity
we denote by ${\cal S}_{\infty}$ and its outward pointing, spacelike
unit normal by $n^{a}$. Starting
with the action (\ref{action}), decomposing ${\cal R}$ into the Ricci
scalar ${}^{(3)}\!{\cal R}$ of the spacelike hypersurfaces and a
divergence term, and integrating by parts several times, gives the VSL
Hamiltonian
\begin{eqnarray}\label{Hamiltonian}
   H=-\int_{\Sigma}d^{3}x\sqrt{g}\left[\frac{e^{\epsilon\phi}}{16\pi G}
   \left({}^{(3)}\!{\cal R}-(\kappa+2\epsilon^{2}) h^{ij}
   \partial_{i}\phi\partial_{j}\phi-2\epsilon\Delta\phi\right)
   +e^{\beta\phi}L_{m}\right]\;\;\;\;\;\;
    \nonumber \\ \;\;\;\;\;\;
   +\frac{1}{8\pi G}\int_{{\cal S}_{\infty}}d^{2}x \sqrt{B\sigma}
   e^{\epsilon\phi}\left(\theta-\epsilon n^{a}\partial_{a}\phi\right),
\end{eqnarray}
where $\theta$ is the extrinsic curvature scalar of, and $\sigma_{ab}$
the induced metric on, the surface ${\cal S}_{\infty}$.
The integral over $\Sigma$ vanishes by virtue of 
Hamilton's equation 
${\delta H}/{\delta \sqrt{B}}=0$, where $\sqrt{B}$ is the lapse
function,   and we are left with the surface
term. In general, this diverges and we must subtract a reference term
$H_{B}$. Following \cite{BY}, we choose $H_{B}$ to have the form
\be
   H_{B}=\frac{1}{8\pi}\int_{{\cal S}_{\infty}}d^{2}x
   \sqrt{B\sigma_{B}}\;e^{\epsilon\phi_{B}}\theta_{B},
\ee
where the subscript ``$B$'' denotes quantities evaluated when $\Sigma$
is intrinsically flat. The Hamiltonian is then
\be\label{Hamiltonian2}
   H=\frac{1}{8\pi}\int_{{\cal S}_{\infty}}d^{2}x
   \sqrt{B\sigma}\;e^{\epsilon\phi}\left(\theta
   -\epsilon n^{a}\partial_{a}\phi\right)-H_{B}.
\ee
We assume that $\phi$ is constant over ${\cal S}_{\infty}$ and denote
its value by $\si$. We further assume the $\phi_{B}=\si$.
One can then show that eqn (\ref{Hamiltonian2}) reduces to 
\be\label{Energy}
   H=e^{\epsilon\phi_{\infty}}\left(M_{ADM}-\half\epsilon Q_{S}\right),
\ee
where 
\be
   Q_{S}:=\frac{1}{4\pi}\int_{{\cal S}_{\infty}} n^{a}\partial_{a}\phi
\ee
is the scalar charge associated with $\phi$. In a spherically
symmetric spacetime, 
\be
   Q_{S}=\lim_{r\to\infty}\left(r^{2}\frac{d\phi}{dr}\right).
\ee
As one would expect from the form of the VSL action, eqn
(\ref{Energy}) is similar to eqn (\ref{MTdef}).

There is an ambiguity in the derivation of eqn (\ref{Energy}) in that
it is not immediately clear if one should multiply $H$ by an additional
power of $e^{\si}$ to account for the effect on the total energy
of varying the value of $c$ throughout a solution. This is related to
the problem discussed in \cite{Mag1} of deciding on the power of
$c_{\infty}$ one must multiply the action (\ref{action}) by.
To show that in
fact eqn (\ref{Energy}) needs no such additional factor we consider
the energy of a star in the weak field limit. Combining the second
equality of eqn (\ref{Lagdef}) with the definition of pressure 
$p:=mc^{2}n^{2}\partial \Pi/\partial n$, one can show that
\be
   (\rho+p)n^{\prime}=np^{\prime}.
\ee
Combining this with eqn (\ref{dpdr}) and integrating gives
\be\label{nrel}
   mc^{2}n=(\rho+p)\sqrt{B}\;e^{\beta\phi}.
\ee
We can define a ``rest energy'' $M_{R}=mc^{2}N$ for the star, which
is simply the sum of the energies of its constituent
particles and is its total energy in the weak field limit.
Combining eqns (\ref{Number}) and (\ref{nrel}) we
have
\be\label{MR}
   M_{R}=\int_{0}^{\infty}4\pi r^{2}\sqrt{B}\;A^{3/2}e^{\beta \phi}
   (\rho+p)\;dr.
\ee
In the lowest order weak field limit, $\rho>>p$, $g_{ab}\ra
\eta_{ab}$ and $\phi$ is constant. In this case, eqns
(\ref{genEin}) and (\ref{MassInt}) together imply that the weak
field ADM mass is
\be
   M_{ADM}=\int_{0}^{\infty}4\pi r^{2}\rho e^{(\beta-\epsilon)\si}\;dr,
\ee
while the Hamiltonian in this limit is simply $e^{\epsilon\si}M_{ADM}$. 
Under the lowest order weak field approximation, this is identical to
the quantity defined by eqn (\ref{MR}) indicating that no additional
powers of $e^{\si}$ need be included in the energy definition and so
we shall adopt eqn (\ref{Energy}) as the definition of the energy of
the star. Due to this expression's similarity with eqn
(\ref{MTdef}), we shall henceforth denote the energy of 
VSL solutions by $M_{T}$.

We may check that our definitions of energy and particle number,
defined by eqn (\ref{Number}) are consistent as follows.
A necessary condition for the stability of stellar solutions in
any theory of gravity is that, given a set of solutions parameterised
by a smoothly varying quantity such as their central density, the
extremal values of the total energy and the conserved particle number
should coincide \cite{KMS}. Using a method similar to the one outlined in
\cite{Jetzer} for boson-fermion stars, it is relatively easy to 
prove that $N$ defined by eqn (\ref{Number}) and our
choice of eqn (\ref{Energy}) for the energy
of a solution are consistent with this requirement in that these two
quantities do have coinciding extrema.

In a metric theory, $M_{P}$ defined by eqn (\ref{MPdef}) 
is also the Kepler mass $M_{K}$ of the star. However,
this is not the case here: the Kepler mass in a VSL theory includes a
term proportional to $\beta$ which embodies the non-minimal coupling
between $L_{m}$ and the metric and, in terms of the variables we are
using here, is given by \cite{Mag2}
\be\label{MKdef}
   M_{K}=M_{P}+\beta Q_{S},
\ee
where, in terms of $M_{ADM}$ and $Q_{S}$, the Poisson mass is
\be\label{MKdef2}
   M_{P}=M_{ADM}-\epsilon Q_{S}.
\ee
These definitions are similar to those in BD theory except that in the
later case, $M_{P}$ and $M_{K}$ are identical.  

%%%%%%%%%%%%%%%%%%%%%%%%%%%%%%%%%%%%%%%%%%%%%%%%%%%%%%%%%
\section{Numerical Results}\label{NUM}
%%%%%%%%%%%%%%%%%%%%%%%%%%%%%%%%%%%%%%%%%%%%%%%%%%%%%%%%%

Equations (\ref{ddAdrr}) to (\ref{dpdr}) are invariant under the
rescaling 
\be\label{rescale1}
   \phi\ra\phi+s,\;\;\;\;A\ra Ae^{(\epsilon-\beta)s}.
\ee
We use this property to set $\phi_{\infty}=0$ for all solutions. In
addition, eqns (\ref{ddAdrr}) to (\ref{dpdr}) are invariant under the
rescaling
\be\label{rescale2}
   A\ra a^{2} A,\;\;\;\;r\ra \frac{r}{a},\;\;\;\;
   Z\ra a Z
\ee
and we use this to set $A_{\infty}=1$ for all solutions. Note that
this last rescaling implies that $\mu\ra\mu/a$.
The function $Z$ automatically satisfies the
boundary condition $Z_{\infty}=0$. We calculate $\mu$, $C$ and
$\lambda$ for each solution using eqns (\ref{mucalc}) to
(\ref{ccalc}), in which we evaluate $A$, $Z$, $\phi$ and their
derivatives at the outermost point of the interior solution.
Equation (\ref{lambda}) is then used to check the
consistency of the numerical calculations. 
We characterise the size of each solution by the Schwarzschild radius $R$,
which is related to $r$ by 
\be\label{Radcalc}
   R:=\sqrt{A_{s}}\;r_{s},
\ee
where the subscript $s$ denotes quantities evaluated at the surface of
the star. This is the same definition of radius used in \cite{Sal}.

Figure 1 shows curves of mass against
central density for several sets of solutions with EOS \ref{EOS1}.
For each set, the
value of $\epsilon$ is fixed, we impose the condition $\epsilon-\beta=4$
and choose the minimum value of $\kappa$ consistent with
eqn (\ref{ppnpars2}). The mass of each solution is calculated using 
eqn (\ref{Energy}). The $\epsilon=4$ solutions are indistinguishable
from those in GR since they are identical to solutions in BD
theory with a large value of $\omega$. The Figure shows that as
$\epsilon$ and $\kappa$ decrease, the energy of all of the solutions
decreases. Masses $M_{T}$ and $M_{K}$, along with other data from the
same sets of solutions, are shown in Table 
\ref{table1}; we do not show values of $M_{P}$ since they are very
close to corresponding values of $M_{T}$.
The data in the table indicate that,
as well as the mass, the radii of the solutions also decreases rapidly
with $\epsilon$.

\begin{table}[ht]
\centering
\begin{tabular}{|c|c|c|c|c|c|c|}\hline 
$\epsilon$ & $\kappa$ & $\kappa/\epsilon^{2}$ & $M_{T}$ & $M_{K}$ &
$Q_{S}$ & $R$ \\ \hline
   4.0    & 53290 & 3330                & 0.711 & 0.712 & -0.0000 & 10.1 \\
   1.0    & 13330 & 13330               & 0.699 & 0.712 & -0.0036 & 10.0 \\
   0.1    & 1333  & $1.33\times 10^{5}$ & 0.588 & 0.711 & -0.0311 & 9.50 \\
   0.005  & 666   & $2.67\times 10^{5}$ & 0.495 & 0.704 & -0.0524 & 8.99 \\
   0.001  & 13.3  & $1.33\times 10^{7}$ & 0.192 & 0.598 & -0.1017 & 6.62 \\
   0.0005 & 6.66  & $2.67\times 10^{7}$ & 0.096 & 0.502 & -0.1016 & 5.27 \\
   0.0001 & 1.33  & $1.33\times 10^{8}$ & 0.012 & 0.269 & -0.0642 & 2.66 \\
\hline
\end{tabular}
\caption{\label{table1}VSL neutron stars with $\epsilon-\beta=4$ and
central density $\rho_{0}=2.9\times 10^{18}$kg$\,$m$^{-3}$. 
The energy $M_{T}$ is calculated using eqn
(\ref{Energy}), the Kepler mass $M_{K}$ using eqn (\ref{MKdef})
and $Q_{S}$ is the scalar charge associated with $\phi$.
All three are measured in units of solar
mass. The radius, in units of km, is calculated using eqn
(\ref{Radcalc}).} 
\end{table}     

The response of the properties of the VSL stars to changes in the
scalar field coupling strength is in marked contrast to to that for
neutron stars in BD theories, data for which are
shown in Table \ref{table2}. (Further
data on the structure of BD stars can be found in Salmona \cite{Sal} but note
that the mass used in this reference is the Poisson mass and the
boundary value of $\Phi$ is different from our value of
$\Phi_{\infty}=1$.)  Values of $\omega$
have been chosen in the Table to match those of $\kappa$ in Table
\ref{table1} and we show masses $M_{T}$ and $M_{K}$; the Poisson mass
$M_{P}$ is identical to $M_{K}$.
The data show that increasing the strength of the
scalar field's coupling has a moderate effect in decreasing the energy
and radius of the stars, while increasing the Poisson mass. 

\begin{table}[ht]
\centering
\begin{tabular}{|c|c|c|c|c|}\hline &&&& \\
$\omega$ & $M_{T}$ & $M_{P}$ & ${\widetilde Q}_{S}$ & 
$R$ \\ \hline
   1333    & 0.711 & 0.711 & -0.0004 & 10.1  \\ 
   666     & 0.711 & 0.712 & -0.0008 & 10.1  \\ 
   13.3    & 0.705 & 0.723 & -0.0365 & 10.1  \\ 
   6.66    & 0.699 & 0.732 & -0.0655 & 10.0  \\
   1.33    & 0.673 & 0.764 & -0.1818 & 9.95  \\
\hline
\end{tabular}
\caption{\label{table2}Neutron stars on BD theory 
with central density $\rho_{0}=2.9\times 10^{18}$kg$\,$m$^{-3}$.
The energy $M_{T}$ is calculated using eqn
(\ref{MTdef}), the Poisson mass $M_{P}$ using eqn (\ref{MPdef})
and ${\widetilde Q}_{S}$ is the scalar charge associated
with the BD scalar field $\Phi$. 
All three are measured in units of solar
mass. The radius, in units of km, is calculated using eqn
(\ref{Radcalc}). For BD theory, the Poisson and Kepler masses are 
identical.} 
\end{table}     

There are two reasons for the differences between BD and VSL neutron
 stars' properties.
From eqn (\ref{ddSdrr}), for small values of $\epsilon$, the 
strength of the coupling between
$\phi$ and the matter varies as $(\kappa/\epsilon)^{-1}$. This quantity
increases as $\kappa$ decreases, which means that the $\phi$ field
becomes more inhomogeneous and the magnitude of $Q_{S}$ increases as
we decrease $\epsilon$. This behaviour is very similar to that for BD
theory. However, in contrast with BD theory, the ADM mass, Poisson mass
and energy $M_{T}$ differ by multiples of  $\epsilon Q_{S}$ and the
product of these two quantities decreases as $\epsilon$ decreases. This
is in contrast with BD theory, where the larger value of ${\wt Q}_{S}$
serves to increase the Poisson (or Kepler) mass to a value significantly
larger that other masses. 

The second, and more significant, cause of structure difference
between the two theories is due to the
non-universal coupling between the matter and the metric in the VSL action.
From eqn (\ref{dpdr}), the pressure gradient has a
term involving the product $\beta\phi^{\prime}$. For all of
these solutions, $\phi^{\prime}<0$ and, under the
assumption that $\epsilon-\beta=4$, $\beta<0$. Hence the product
$\beta\phi^{\prime}>0$ and since $p$ must be a decreasing function of
$r$, eqn (\ref{dpdr}) implies that the term involving $\beta$
increases the magnitude of the pressure gradient, leading to $p\ra 0$
at a smaller value of $R$ than for a star in metric theory with an
identical central density. The significantly
smaller physical size of the star means that its energy is also much
smaller. 

Figure 2 shows mass against central density for sets of solutions with EOS
\ref{EOS2}, with the same limits on the parameters $\epsilon$ and
$\beta$ as for the solutions of Figure 1. These solutions show the
same overall behaviour as those with the simpler EOS: as the parameter
$\eta$ decreases, the solutions become less massive for a given
central density. As before, the $\epsilon=4$ solutions are
indistinguishable from those of GR.
In addition, the central density at which the maximal
mass solution occurs increases as $\epsilon$ decreases, again 
in a similar way to the solution shown in Figure 1. In addition
we have found that, for a given central density, the radius of a
solution decreases as $\epsilon$ decreases. However, both this feature
and the reduction in mass are less severe than for the simpler EOS and
both maximal mass and minimum radius of the maximal mass solution
decrease far less rapidly with decreasing $\epsilon$. This is due to
the difference in the stiffness between the two equations of state.

Since the Figure 2 solutions have a reasonably realistic EOS, we can
use them to place a limit on the 
parameters of the VSL theory for the case
studied here in which $\epsilon-\beta=4$. The masses of neutron stars
in binary systems detected to date have masses in the range $(1.36\pm
0.8)$ solar masses, while neutron stars in orbit about other types of
companion body have masses of at least 1.4 solar masses (for a review
of this data and references, see \cite{Heis}). Hence, for a VSL theory
to be viable, it must allow for the existence of neutron stars whose
mass is at least 1.3 solar masses. From Figure 2, 
this implies that the parameter $\epsilon$ must satisfy the 
inequality $\epsilon\geq 0.00015$. 

%%%%%%%%%%%%%%%%%%%%%%%%%%%%%%%%%%%%%%%%%%%%%%%%%%%%%%%%%
\section{Sensitivity and Evolution of Mass}\label{SENS}
%%%%%%%%%%%%%%%%%%%%%%%%%%%%%%%%%%%%%%%%%%%%%%%%%%%%%%%%%

For many theories of gravity, such as BD theory and other more
general scalar-tensor (ST) theories, solar system observations place strong
constraints on the theories' parameters via the PPN formalism, while 
limits derived from observation of strong field phenomena are far
weaker. However, for VSL
theories with small values of $\epsilon$, the PPN limit (\ref{ppnpars2})
on $\kappa$ is
extremely weak and we should therefore look to strong field
observations, in particular binary pulsar data, to provide the
strongest constraints. The situation here is similar to that found
by Damour and Esposito-Farese for certain classes of ST theory which
satisfy PPN limits to an arbitrarily high degree but show significant
strong field effects \cite{DEF} and references therein. 
We shall leave the full treatment of the motion of binary pulsars in a
VSL theory to a future publication. Here we shall briefly mention
some of the key ideas involved and use part of the formalism to
discuss how VSL stars will evolve in a cosmological setting.

The analysis of binary pulsar motion is based on the parameterised
post-Keplerian (PK) formalism, which treats each star as a point mass
moving according to quasi-Newtonian equations of motion. Details of
the particular theory under study are made manifest by allowing the
active, passive and inertial masses of each star to 
vary according to the value of any ambient,
non-metric gravitational fields. At the lowest level of approximation,
the variability of a particular mass
is quantified by the sensitivity $s$, given by
\be\label{sdef}   
   s:=\left.\sum_{A}\frac{\partial(\log M)}
   {\partial(\log\psi^{A})}\right|_{\psi^{A}_{\infty}},
\ee
where $M$ is the mass under consideration and the $\psi^{A}$ are the 
non-metric
gravitational fields. In ST theories of gravity, there 
is only one extra field, the
scalar field $\Phi$ and it turns out that active, passive and inertial
masses are proportional to each other. 
For the VSL theory we are considering here,
we expect much of the formalism
to be very similar to that for BD gravity, since the vacuum, 
inter-body field equations are the same for both theories. Since the
BD and VSL scalar fields are related by $\Phi=e^{\epsilon\phi}$, we
shall define $s$ to be
\be\label{VSLsens}
   s=\frac{\partial (\log M)}{\partial (\epsilon\phi)}
   =\frac{1}{\epsilon M}\frac{\partial M}{\partial \phi},
\ee
where again $M$ denotes any mass. 

We shall next derive an expression for the sensitivity associated with
the Hamiltonian, identical to the VSL tensor mass $M_{T}$, of a compact
object in a VSL theory. We consider
first the variation of the Hamiltonian (\ref{Hamiltonian}) 
induced by a variation of
$\phi$ on ${\cal S}_{\infty}$. Taking the functional derivative of
this equation, performing several integration by parts and noting that
the volume term vanishes and the terms involving
$\partial_{a}(\delta\phi)$ in the surface integral cancel, we are left
with
\be
   \delta H=\frac{1}{8\pi G}\int_{{\cal S}_{\infty}}
   d^{2}x\sqrt{B\sigma} e^{\epsilon\phi}\left(\epsilon\theta
   -\epsilon n_{a}U^{b}\nabla_{b}U^{a}
   +\kappa n^{a}\partial_{a}\phi\right)\delta\si+\delta H_{B},
\ee 
where we have assumed that $\delta\phi$ is constant over ${\cal
S}_{\infty}$ and have denoted it by $\delta\si$.
The reference term $H_{B}$ is linear in
$e^{\phi_{\infty}}$. Thus
\be
   \delta H_{B}=\epsilon H_{B}\delta\phi_{\infty}.
\ee
Combining the above two equations with eqn (\ref{Hamiltonian}), we have
\be\label{HH}
   \delta H=\frac{1}{8\pi G}\int_{{\cal S}_{\infty}}
   d^{2}x\sqrt{B\sigma} e^{\epsilon\phi}\left[
   (\kappa+\epsilon^{2})n^{a}\partial_{a}\phi
   -\epsilon n_{a}U^{b}\nabla_{b}U^{a}\right]\delta\si
   +\epsilon H \delta\phi_{\infty}.
\ee  
Using the definition (\ref{MPdef}), one can
show that
\be
   \frac{1}{8\pi G}\int_{{\cal S}_{\infty}}
   d^{2}x\sqrt{B\sigma}\;n_{a}U^{b}\nabla_{b}U^{a}=\half M_{P}.
\ee
In addition, the first term of eqn (\ref{HH})
is simply $\half(\kappa+\epsilon^{2})Q_{S}$. 
Thus we have 
\be
   \delta H=e^{\epsilon\si}\left[\half(\kappa+\epsilon^{2})
   {Q_{S}}-\half\epsilon M_{P}\right]\delta\si+\epsilon H \delta\si.
\ee
Finally, using the relations (\ref{Energy}) and (\ref{MKdef2}), and
taking the limit $\delta\si\ra 0$,
one can show that
\be
   \frac{\partial H}{\partial \si}=\half \epsilon H 
   +e^{\epsilon\si}\left(\frac{3\epsilon^{2}+2\kappa}{4}\right)Q_{S}.
\ee
Hence the sensitivity associated with both $H$ and the identical
quantity $M_{T}$ is
\be\label{Sens}
   s=\half+e^{\epsilon\si}\left(\frac{3\epsilon^{2}+2\kappa}
   {4\epsilon}\right)\frac{Q_{S}}{M_{T}}.
\ee
Due to the similarities between the two theories, we expect $s$
to appear in the PK formalism for a VSL theory in much the same way as
in a ST theory. Thus we expect that, for any particular mass $M$, 
presence of the $\phi$ field introduces a strong field  correction of the
form $M\ra M(1+\nu s)$, where $\nu$ is a parameter which depends the type
of mass considered. Equation (\ref{Sens}) then indicates
how strong field gravitational effects depend upon the values of
$\epsilon$ and $\beta$.

Equation (\ref{Sens}) may be used to investigate the gravitational
evolution of a compact body in a VSL theory.  The scalar field far
from a compact body such as a neutron star should match smoothly to the
cosmological value of the field. Hence, assuming that a body embedded
in a cosmological solution can be modelled approximately by an
isolated, asymptotically flat solution, one expects that
the boundary field $\si$ to follow the cosmological evolution 
of $\phi$. Equation
(\ref{Sens}) then implies that the evolution of the mass of the body
depends strongly on the relative values of $\epsilon$ and $\kappa$.  
For all of the solutions shown in Figure \ref{fig1} and
Table \ref{table1}, except the $\epsilon=4$ solutions, $s<0$. The
primary reason for this is that, as $\epsilon$ decreases, the ratio
$\kappa/\epsilon$ which dominates the second term of eqn (\ref{Sens})
increases dramatically. In a cosmological solution,
$\phi$ should be a decreasing function of cosmological time so as to be
consistent with the requirement that $c$ was larger in the
past. 
Use of eqn (\ref{Sens}) then implies that the energy of these VSL stars
increases with cosmological time, the largest rates of increase
coinciding with smallest values of $\epsilon$. This is in marked
contrast with ST theories for which, in general, the energy of a
compact body decreases with cosmological time \cite{AW,CS}.
 
%%%%%%%%%%%%%%%%%%%%%%%%%%%%%%%%%%%%%%%%%%%%%%%%%%%%%%%%%
\section{Conclusions}\label{CONC}
%%%%%%%%%%%%%%%%%%%%%%%%%%%%%%%%%%%%%%%%%%%%%%%%%%%%%%%%%

We have examined the properties of neutron stars in one particular
VSL theory in which the local value of the speed of light is
determined by a scalar field $\phi$. 
The theory bears a more than superficial
resemblance to BD and other ST theories, and we have found that the
expression giving the energy of an asymptotically flat solution to the
VSL field equations is very similar to it counterpart in ST
theories. However, despite these similarities, the properties of VSL
neutron stars are very different to their counterparts in VSL
theories. In particular, for certain choices of $\epsilon$ and
$\beta$, VSL neutron stars are limited to being much smaller those in
GR and the dependence of mass on the strength of the
coupling between $\phi$ and the matter is extremely strong. 
This allows one to use the mere existence of
neutron stars to place constraints on the theory's parameters, even
without having to analyse the motion of binary pulsars: the maximum
neutron star mass allowed by the theory must obviously be at least as
large as those of observed neutron stars. Unfortunately, neutron star
masses are fairly sensitive to the EOS used (see, for
example, \cite{Diaz,Arnett}), although we have placed a rather crude
limit on the parameter $\epsilon$ using a reasonably realistic
equation of state.

We have also given a brief analysis of the evolution of a VSL star in
a cosmological setting found that, as the cosmological background
evolves and $\phi$ decreases in value, the energy of a VSL star
increases in general. This result is independent of the EOS
used or even of the type of matter of which the star is
composed  and could have serious implications for the
stability of a star since its fractional binding energy $(M_{T}-N)/N$
could eventually become positive.

To properly place limits on the parameters of a VSL theory, one needs
a full analysis of the predicted behaviour of binary pulsar
motion. However, this is not the only area of astrophysics in which a
VSL theory might give significantly different predictions to those of
other theories. For example, it was shown in \cite{Mag2} that black
holes in this theory are very different to their GR counterparts in
that the proper time taken for an in-falling  particle to reach the
horizon can be infinite. If this
is indeed the case, a varying speed of light
may have a significant effect on the physics
of black hole accretion disks since matter orbiting
the black hole would never enter the black hole horizon.

%%%%%%%%%%%%%%%%%%%%%%%%%%%%%%%%%%%%%%%%%%%%%%%%%%%%%%%%%
\subsection*{Acknowledgements}

I would like to thank the University of Sussex for its hospitality
while this work was completed.

%%%%%%%%%%%%%%%%%%%%%%%%%%%%%%%%%%%%%%%%%%%%%%%%%%%%%%%%%

%%%%%%%%%%%%%%%%%%%%%%%%%%%%%%%%%%%%%%%%%%%%%%%%%

\newpage
\begin{figure}
\begin{center}
\includegraphics{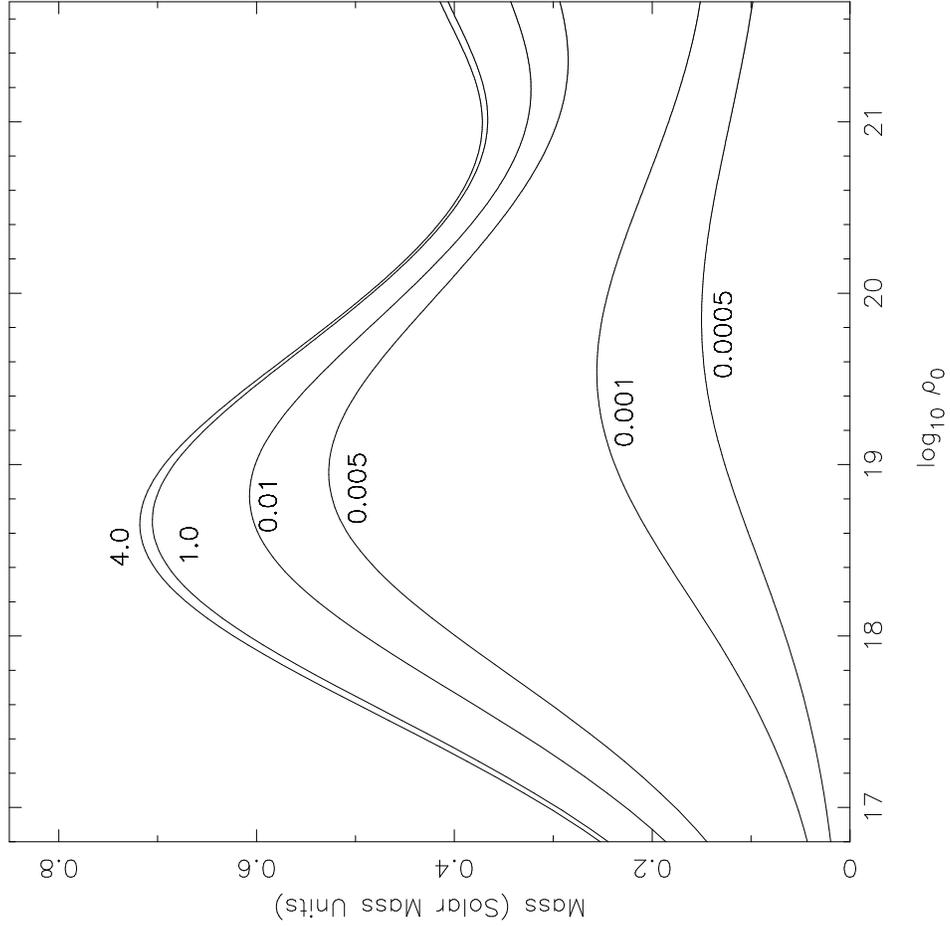}
\caption{\label{fig1}Mass curves for neutron stars in the VSL
theory with the simple, non-interacting Fermion EOS (\ref{EOS1}). 
The curves are labelled by their value of $\epsilon$. For each
curve, $\epsilon-\beta=4$ and the value of $\kappa$ is the minimum
value consistent with the weak field, solar system constraints. The
central density is given in units of kg$\,$m$^{-3}$.}
\end{center}
\end{figure} 

\begin{figure}
\begin{center}
\includegraphics{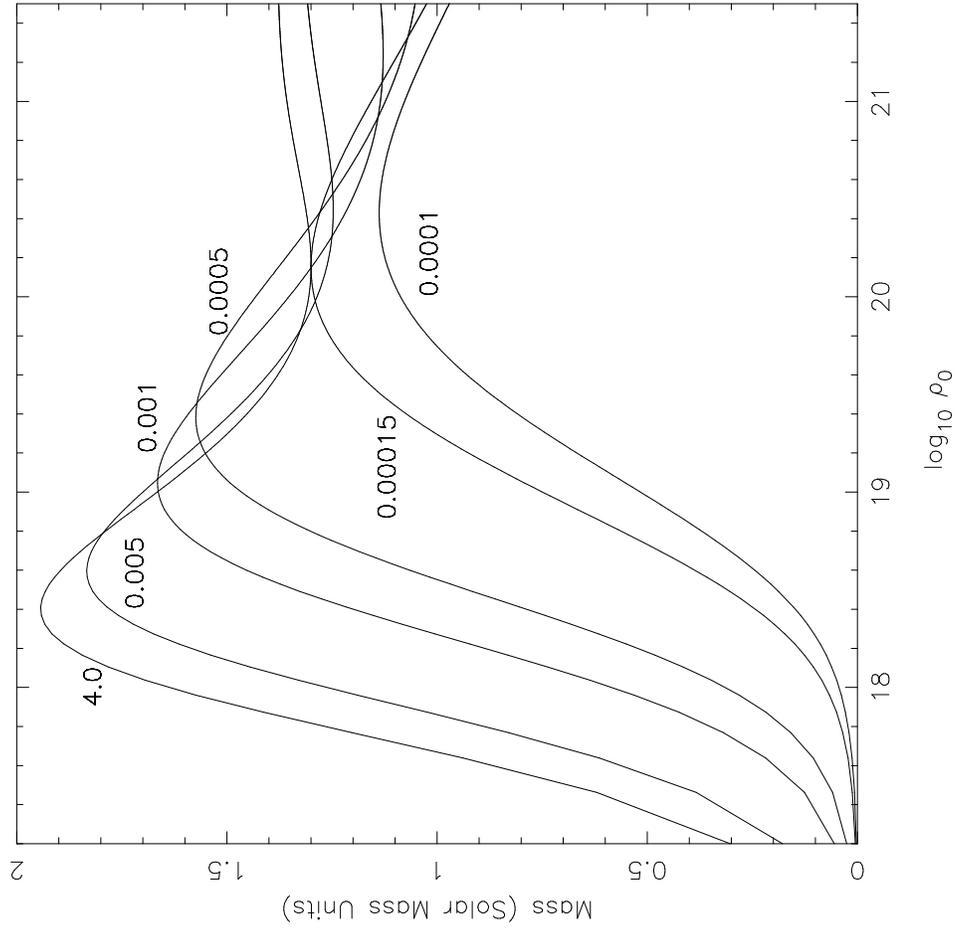}
\caption{\label{fig2}Mass curves for neutron stars in the VSL
theory with the realistic EOS (\ref{EOS2}). Units, labels and curve
parameterisation is the same as in Figure \ref{fig1}.}
\end{center}
\end{figure} 

\end{document}